%% file: draft.tex
\documentclass{article}
\usepackage{spconf,amsmath,graphicx}
\usepackage{booktabs}
\usepackage{url}
\input{math_commands.tex}

\usepackage{multirow}
\usepackage{array}
\usepackage{longtable}
\newcolumntype{C}[1]{>{\centering\arraybackslash}m{#1}}
\title{DurIAN-E: Duration Informed Attention Network For Expressive Text-to-Speech Synthesis}
\name{Yu Gu, Yianrao Bian, Guangzhi Lei, Chao Weng, Dan Su}
\address{Tencent AI Lab}
\begin{document}
%\ninept
%
\maketitle
\begin{abstract}
This paper introduces an improved  duration informed attention neural network (DurIAN-E) for expressive and high-fidelity  text-to-speech  (TTS) synthesis.  Inherited from the original DurIAN model, an auto-regressive model structure in which the alignments between the input linguistic information and the output acoustic features are inferred from a duration model is adopted. Meanwhile the proposed DurIAN-E utilizes multiple stacked SwishRNN-based Transformer blocks as linguistic encoders.
Style-Adaptive Instance Normalization (SAIN) layers are exploited into 
frame-level encoders to improve the modeling ability of expressiveness.
A  denoiser incorporating both  denoising diffusion probabilistic model (DDPM) for mel-spectrograms and SAIN modules is conducted to further improve the synthetic  speech quality and expressiveness.
Experimental results prove that the proposed expressive TTS model in this paper can achieve better performance than the state-of-the-art approaches  in both subjective mean opinion score (MOS) and preference tests.
\end{abstract}
\begin{keywords}
Expressive TTS, DurIAN, SwishRNN, Transformer, Style-Adaptive Instance Normalization, DDPM
\end{keywords}
\section{Introduction}
\label{sec:intro}

Text-to-speech (TTS) synthesis is the task of generating intelligible and natural sounding synthetic speech waveforms given the input text messages. 
TTS synthesis technique is an indispensable basic component in various applications with speech interface such as car navigation systems, voice assistant and screen readers, etc. 
Due to the advantages of deep learning,  many state-of-the-art TTS systems  based on deep neural networks were able to synthesize more natural and high-quality speech,
compared with traditional unit selection  concatenative and statistical parametric speech synthesis approaches.   Those  different kinds of acoustic models and neural vocoders can be divided into autoregressive (AR) and  non-autoregressive methods. Some sequence-to-sequence acoustic models relied on content-based attention mechanism to address the one-to-many alignment problems, which could generate mel-spectrograms  from linguistic features frame-by-frame using AR decoders \cite{tacotron2}. Different explicit phoneme duration models rather than attention modules were also employed into some non-autoregressive TTS acoustic models in which acoustic feature sequences of each utterance could be generated in parallel \cite{renfastspeech,  paralleltacotron2}. The DurIAN model \cite{yu2020durian} also involved an additional duration model to avoid the typical  attention errors such as skipping and repeating and meanwhile the AR decoders were also reserved to improve speech quality by combining 
both the linguistic information and the acoustic information from previously predicted acoustic features.

Although those TTS systems have synthesized speech qualitatively similar to real human speech, there still exists a huge gap between TTS-synthetic speech and human speech in terms of expressiveness. Many researchers have also  focused on expressive TTS technology for decades, which aims to model and control the speaking style
and can further broaden TTS application prospect. At present, there are 
two mainstream approaches to model the speaking style information:
one uses pre-defined categorical style labels as the global control condition of TTS systems to denote different speaking styles \cite{tits2019visualization, tits2020exploring} and the other imitates the speaking style given a reference speech \cite{wang2018style, skerry2018towards}. For the first kind of approach, the style control strategy is more intuitive and interpretable, which is more suitable for practical applications. For the second one, the global style tokens or style embeddings extracted from the training datasets can enrich the diversity of expressiveness and additional style labels is not required.

Some style adaptive and transfer methods were also conducted in expressive TTS systems. Style-Adaptive Layer Normalization (SALN) layers were applied on expressive TTS systems, which received the style representation vector and predicted the gain and bias of the input feature vector \cite{min2021meta, liu2021meta}.
Style-Adaptive Instance Normalization (SAIN) layers \cite{li2022styletts} were also employed to learn the distribution with a style-specific mean and variance for each channel in mel-spectrograms  and  each channel in the mel-spectrograms represents a single frequency range. Therefore compared with SALN, where a single mean and variance was learned for the entire feature map,  decoders with SAIN blocks achieved better expressiveness in terms of style reflection than SALN blocks \cite{li2022styletts}. SwishRNN \cite{lei2022simple},  an extremely simple recurrent module, was built into the Transformer-based masked language model to increase model stability and accuracy.
Recently denoising diffusion models \cite{ddpm} were also applied as acoustic models of TTS systems, which converted the noise into mel-spectrogram conditioned on the linguistic features and could achieve better speech quality \cite{jeong2021diff,liu2022diffsinger}. 
Motivated by the success of SwishRNN-based Transformer, a linguistic encoder is proposed in the 
DurIAN-E model, in which SwishRNNs are involved to substitute the feed-forward blocks in Transformer \cite{transformer}. SAIN-based modules are employed on both frame-level encoders
and DDPM-based denoiser to better model the expressiveness with the pre-defined rich categorical style labels.  An AR decoder similar-like the original DurIAN model is  reserved to
take full advantages of both style-specific linguistic features and acoustic features.

This paper is organized as follows. Section \ref{sec:related} gives a brief review of the  modules and models related with our work.
Section \ref{sec:duriane} introduces the proposed  DurIAN-E model in this paper and the constructed expressive TTS system in detail. The experimental conditions and results are described in  Section \ref{sec:exp} and finally Section \ref{sec:con}  concludes this paper.

\section{Related Work}
\label{sec:related}
\subsection{DurIAN}
\label{subsec:durian}
DurIAN \cite{yu2020durian} is an AR model in which the alignments between the
input text and the output acoustic features are inferred from a duration model, which is different from the conventional end-to-end attention mechanism used in speech synthesis systems such as Tacotrons \cite{tacotron2}. The architecture of DurIAN mainly contains a \textit{skip-encoder}  to encode the phoneme and prosody sequences,  a \textit{duration-model} which aligns the input phoneme sequence and the target acoustic frames at frame level, an AR decoder network that generated target acoustic features frame by frame and a \textit{post-net} \cite{tacotron2} to further enhance the quality of the predicted mel-spectrograms.

The \textit{skip-encoder} utilizes a sequence of symbols $\{\vx[i]\}_{i=1}^N$ as input
 which contains both the phoneme sequence and the prosodic boundaries among different phonemes.
The output hidden state sequences are encoded as $\{\vh[i]\}_{i=1}^{N'},$ 
where $N$ is the length of the input sequence, and $N'$ is the length of the input phoneme sequence without the prosodic boundaries. 
It's worth noting that 
the length $N'$  is smaller than the length $N$ of the input sequence because the hidden states associated with the prosodic boundaries are excluded by \textit{skip-encoder}. 
Then according to the frame numbers predicted from the \textit{duration-model},  the sequence $\{\vh[i]\}_{i=1}^{N'}$ is expanded by replication as frame aligned hidden states $\{\ve[i]\}_{i=1}^{T}$ where $T$  is the sum number of acoustic frames.  During the training stage, the ground truth duration of each phoneme is obtained through forced alignment using GMM-HMMs. 
The duration model is jointly trained conditioned on $\{\vh[i]\}_{i=1}^{N'}$  to minimize the $\mathrm{\ell2}$ loss between the predicted and target duration obtained from forced alignment.

Similar with other end-to-end models such as Tacotrons, the expanded hidden states $\{\ve[i]\}_{i=1}^{T}$  which are exactly paired with the target acoustic frames are employed as the AR decoder to predict each mel-spectrogram frame autoregressively.
Then the output acoustic features from the decoder network is passed through a \textit{post-net} \cite{tacotron2} with convolutional layers and residual connections to further improve the quality of the predicted mel-spectrograms.
The entire acoustic models are trained to minimize the $\mathrm{\ell1}$ loss.

\subsection{SwishRNN}
\begin{figure}[t]
  \centering
  \includegraphics[width=.9\linewidth]{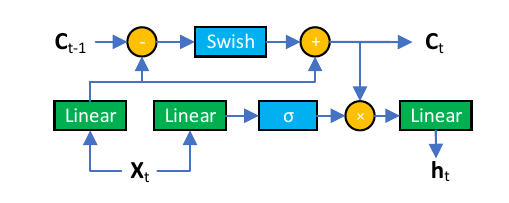}
  \caption{The SwishRNN cell.}
  \label{fig:swishrnn}
\end{figure}

 SwishRNN \cite{lei2022simple} consists of a multiplicative gating  recurrent cell which uses only two matrix multiplications and an extremely simple sequential pooling operation. Therefore SwishRNN is much faster than other heavier RNNs such as LSTM and GRU. As illustrated in Fig.\ref{fig:swishrnn}, SwishRNN first conducts two linear transformations of input sequence $\mX$:
\begin{align}
    \{\vx_1[i]\}_{i=1}^l = \mX \mW_1, \quad   \{\vx_2[i]\}_{i=1}^l = \mX \mW_2
\end{align}
where $\mW_1$ and $\mW_2$ are  parameter matrices optimized during training and $l$ is the sequence length.
The hidden vectors $\{\vc[i]\}_{i=1}^l$ are calculated as follows:
\begin{align}
\label{equ:re}
    \vc[i] = \texttt{Swish}\left(\vc[i\text{-}1] - \vx_1[i]\right) + \vx_1[i],
\end{align}
where $\texttt{Swish}()$ represents the element-wise Swish activation function~\cite{ramachandran2017searching}.\footnote{$\texttt{Swish}(\vx) = \texttt{sigmoid}(\alpha\cdot \vx + \beta) \cdot \vx$.}
Eq.(\ref{equ:re})  can be interpreted as a pooling operator where the greater value between $\vc[i\text{-}1]$ and $\vx_1[i]$ are selected.\footnote{Note $\vc[i]=\vx_1[i]$ if $\vx_1[i] \gg \vc[i\text{-}1]$, and $\vc[i]=\mathbf{c}[i\text{-}1]$ if $\vx_1[i] \ll \vc[i\text{-}1]$.}
Finally, the output sequences are calculated followed by a linear layer with weight $\mW_3$:
\begin{align}
%  \mathbf{H} = \mW_3 \left ( \mathbf{C}\odot \sigma(\mathbf{X}_2) \right ) + \vb_3
\mathbf{H} = \mW_3 \left ( (\mathbf{C} + \vb_c) \odot \sigma(\mX_2 + \vb_\sigma) \right ) + \vb_3,
\end{align}
where $\sigma()$ is a \texttt{sigmoid} gating activation function, $\odot$ is the  element-wise product, $\mC$ and $\mX_2$ represent the concatenated matrices of $\{\vc[i]\}_{i=1}^l$ and  $\{\vx_2[i]\}_{i=1}^l$ respectively.
\begin{figure}[t]
  \centering
  \includegraphics[width=\linewidth]{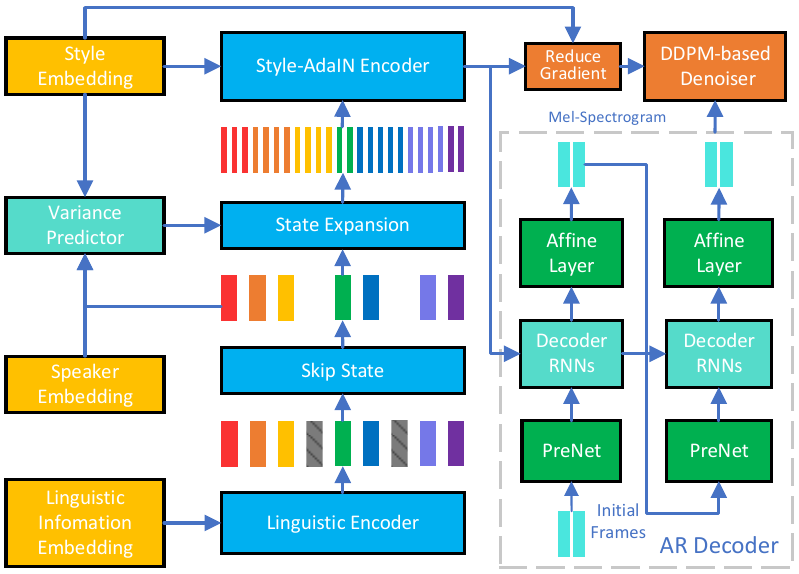}
  \caption{Model structure of DurIAN-E.}
  \label{fig:structure}
\end{figure}

\subsection{Denoising diffusion probabilistic model}
DDPMs take inspiration from non-equilibrium statistical physics in which the main idea is to  iteratively destroy the structure in data through a  \textit{ diffusion process}, and afterward,
to learn a  \textit{reverse process} to restore the data structure.
The \textit{diffusion process} is modeled as a Gaussian transformation chain from data $\vx_0$ to the latent variable $\vx_T$ with pre-defined variance schedule $\beta_1, \cdots,\beta_T$:
\begin{equation}
    q(\vx_{1:T}|\vx_0)=\prod_{t\geq 1}q(\vx_t|\vx_{t\!-\!1}),
\end{equation}
where $q(\vx_t|\vx_{t\!-\!1})\sim\mathcal{N}(\vx_t;\sqrt{1-\beta_t}\vx_{t\!-\!1},\beta_t\mathbf{I})$, $T$ is the total iteration step and $q(\vx_0)$ is the original data distribution .
The  \textit{ reverse process} parameterized with $\theta$ is a denoising function to remove the added noise to restore the data structure, which is defined by:
\begin{equation}
    p_\theta(\vx_{0:T})=p(\vx_T)\prod_{t\geq 1}p_\theta(\vx_{t\!-\!1}|\vx_t).
\end{equation}
The denoising distribution $p_\theta(\vx_{t\!-\!1}|\vx_t)$ is often modeled by a conditional Gaussian distribution as $p_\theta(\vx_{t\!-\!1}|\vx_t)\sim\mathcal{N}(\vx_{t\!-\!1};\mathbf{\mu}_\theta(\vx_t), \sigma_t^2\mathbf{I})$, where $\mu_\theta(\vx_t, t)$ and $\sigma_t^2\mathbf{I}$ are the corresponding mean and variance.
Through the parameterized reverse process with the well-trained parameter $\theta$, 
the target data $\vx_{0}$ can be sampled from a Gaussian noise $\vx_T \sim \mathcal{N}(\mathbf{0}, \mathbf{I})$ iteratively for $t=T, T\!-\!1, \cdots, 1$, in which $\vx_{t\!-\!1}$ is sampled following distribution $p_\theta(\vx_{t\!-\!1}|\vx_t)$.
The training  goal is to maximize the evidence lower bound (ELBO$\leq\log p_\theta(\vx_0)$), which can be optimized to match the true denoising distribution $q(\vx_{t\!-\!1}|\vx_t)$ with the parameterized denoising model $p_\theta(\vx_{t\!-\!1}|\vx_t)$ with:
\begin{equation} 
\label{eq4}
    \text{ELBO} = \sum_{t\geq 1}\mathbb{E}_{q_({\vx_t})}[D_{KL}(q(\vx_{t\!-\!1}|\vx_t)||p_\theta(\vx_{t\!-\!1}|\mathbf{x}_t)],
\end{equation}
where $D_{KL}$ denotes the Kullback-Leibler (KL) divergence. Then Eq.\ref{eq4} can be further transformed and reparameterized as  a simple regression problem to optimize the MSE loss between the sampled and predicted noise terms \cite{ddpm}.

\section{DurIAN-E}
\label{sec:duriane}
\subsection{Architecture}

The model structure of the proposed DurIAN-E is depicted in Fig. \ref{fig:structure}. DurIAN-E basically adopts the  original 
 model architecture of DurIAN.
The \textit{skip-encoder} mechanism and AR decoder networks are reserved and auxiliary duration information is also 
involved to reduce  word skipping/repeating errors. The variance predictors \cite{renfastspeech} which include phoneme-level duration, pitch and pitch range predictors are employed to 
add enough variance information to the hidden sequence and improve the prosody modeling ability. 
The ground-truth values of duration, pitch and pitch range
extracted from the recordings are used as inputs into the hidden sequence in the training stage and the predicted values are utilized in the inference stage. 
 At the same time, those ground-truth values are also used as targets to train the variance predictor. Style and speaker embeddings are also added to the hidden input sequence to better distinguish different utterances from multiple styles and speakers.

Different from the original DurIAN, we split the encoder into a phoneme-level linguistic encoder using SwishRNN-based Transformers and a frame-level encoder combining SAIN to improve expressiveness.
Meanwhile motivated by diffusion model based TTS models \cite{liu2022diffsinger}, the \textit{post-net} is replaced by a style controllable DDPM-based denoiser to achieve better speech quality. Section \ref{subsec:encoders} and \ref{subsec:denoiser} introduce these modules in detail.

\begin{figure}[t]
  \centering
  \includegraphics[width=\linewidth]{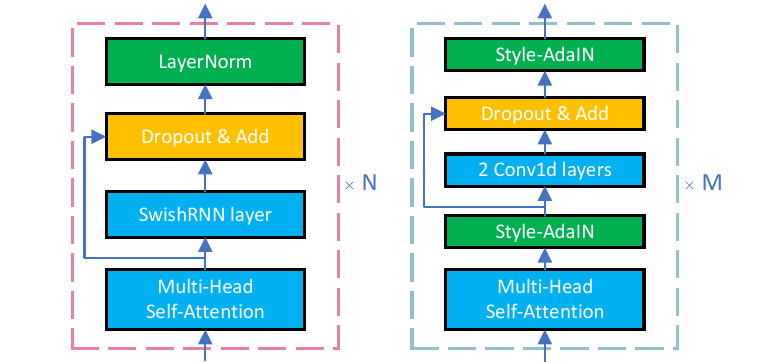}
  \caption{Blocks in the encoders of DurIAN-E.}
  \label{fig:block}
\end{figure}

\subsection{Encoders}
\label{subsec:encoders}
\subsubsection{Linguistic encoder}
The linguistic encoder is a phoneme-level model and only employs linguistic features as input. As described in Section \ref{subsec:durian}, the linguistic information sequence includes the phoneme sequence and the prosodic boundaries among different phonemes. The linguistic encoder is conducted by 4 Transformer blocks with hidden size of 256 and each block interleaves a two-headed attention module, a SwishRNN layer, residual connection and layer normalization shown in left side of Fig. \ref{fig:block}. 
Comparing with the FFN blocks in standard Transformer \cite{transformer}, the recurrent architecture in SwishRNN is more capable of modeling the temporal relationship and order between different phonemes in the linguistic sequences and  meanwhile combining recurrence and attention can also improve the model stability.

\subsubsection{Frame-level encoder}
For the frame-level encoder, the input sequence is expanded using given phoneme duration, of which  the length is equal to that of the target mel-spectrogram and much larger than the phone-level encoder. Due to the longer input sequence, frame-level encoder doesn't follow the recurrent backbones of the linguistic encoder for greater efficiency. Similar with FastSpeech, 4  Transformer blocks with hidden size of 256 using 2 convolutional layers with 
kernel size of 9 rather than SwishRNNs
are employed to encode the expanded intermediate hidden state. As displayed in the right side of Fig. \ref{fig:block}, the layer normalization is substituted by  a SAIN layer to improve the expressiveness and achieve style control. SAIN is defined as following:
 \begin{equation}
    \label{equ:adain}
    \text{Style-AdaIN}(\vx, \vs) = G(\vs) \frac{\vx - \mu(\vx)}{\sigma(\vx)} + B(\vs),
\end{equation} 
where $\vx$ is a single channel of the feature maps,
 $\vs$ is the style embedding,
 $\mu(\cdot)$ and $\sigma(\cdot)$ denote the channel mean and standard deviation, and $G$ and $B$ are learned linear projections for computing the adaptive gain and bias according to the style vector $\vs$. 
 The normalization is operated along the sequence and the style-specific statistics (mean and variance)
 for different channels are independent of each other in every utterance.
 Style vector $\vs$ shares the identical embedding with the variance predictor and DDPM-based denoiser, which is generated from categorical style labels. 

\subsection{DDPM-based denoiser}
\label{subsec:denoiser}
\begin{figure}[t]
  \centering
  \includegraphics[width=\linewidth]{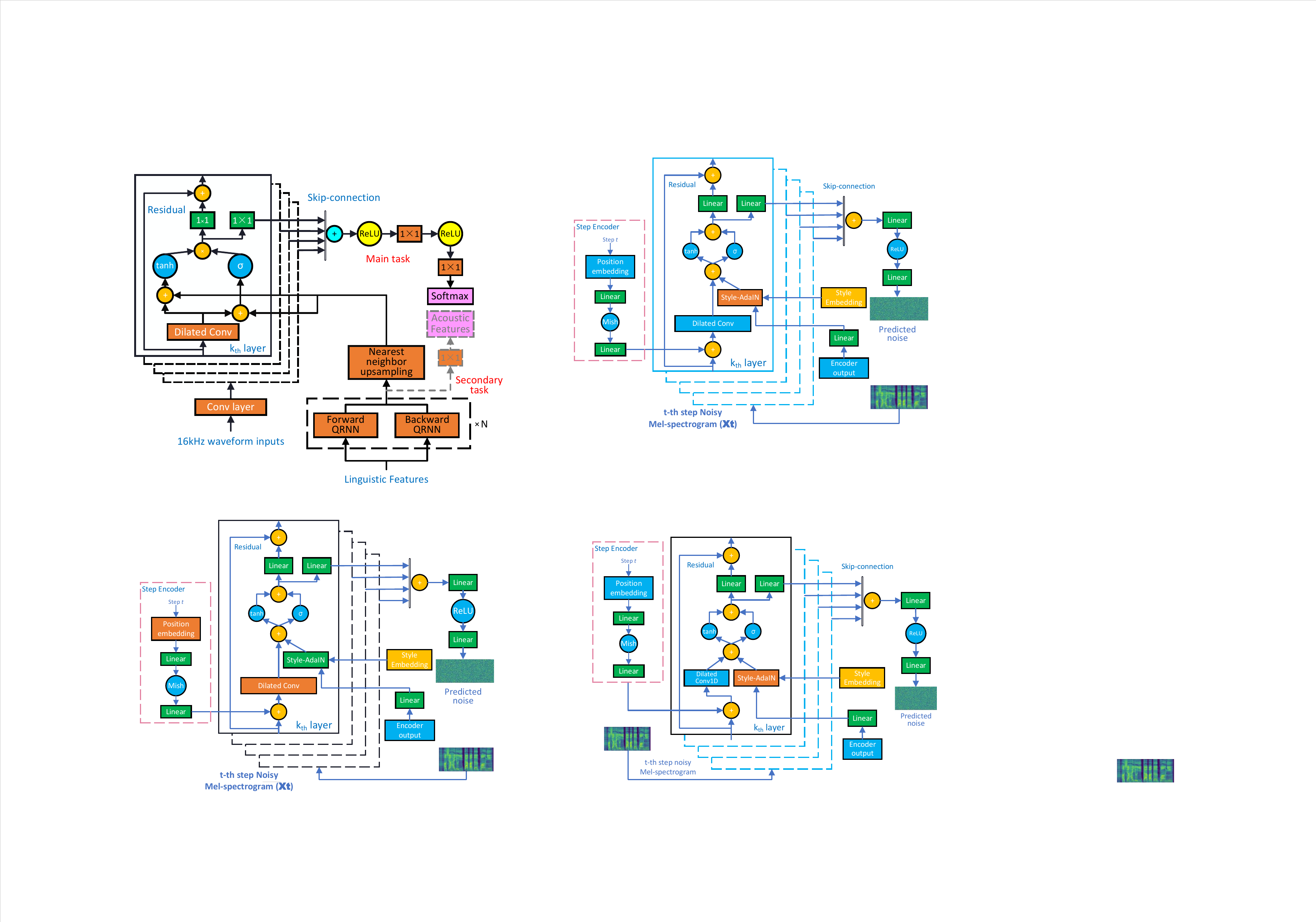}
  \caption{Network structure of the denoiser in DurIAN-E.}
  \label{fig:denoiser}
\end{figure}
As illustrated in Fig. \ref{fig:structure}, the output mel-spectrograms generated frame-by-frame through the AR decoder are delivered into a denoiser to further improve the speech quality. The \textit{post-net} used in DurIAN is
replaced by a DDPM. Similar with other TTS applications using diffusion models \cite{liu2022diffsinger, diffwave},
the non-causal WaveNet \cite{wavenet} architecture is also adopted. As exhibited in Fig. \ref{fig:denoiser}, the denoiser is composed of a stack of 20
residual blocks including a convolutional layer with kernel size of 3, a gated activation unit and element-wise adding operations. The outputs of each block are added together through the skip-connection 
 to predict the noise term  $\bepsilon_\theta(\cdot)$ at $t$-th step. A step encoder is established to  make a distinction between different steps. The denoiser is conditioned on the output sequence of the frame-level encoder. The SAIN layer in Eq. \ref{equ:adain} is also deployed on each residual block to enhance the effect of different style labels.

The shallow diffusion mechanism proposed in DiffSinger \cite{liu2022diffsinger} is also adopted in DurIAN-E. For the training stage, the loss for the denoiser is calculated as 
\begin{align}
\label{eq:loss}
    \Eb{\vm_0, \bepsilon}{ \lambda_{t} \left\| \bepsilon - \bepsilon_\theta(\sqrt{\bar\alpha_t} \vm_0, \sqrt{1-\bar\alpha_t}\bepsilon, \vs, \vc,t) \right\|^2},
\end{align}
where $\vm_0$ is the grouth truth mel-spectrogram, $\bepsilon$ is the gaussian noise, $\vs$ is the style embedding, $\vc$ denotes the output of frame-level encoder, $\lambda_t$ and $\bar\alpha_t$ is the pre-defined coefficients corresponding to the step index $t$. $t$ is randomly sampled from 1 to the total step $T$ for every training step. We decrease the gradient values from the denoiser back-propagated to other parts of the model by a factor of 10 so that influence of the auxiliary denoiser to other modules can be reduced. For the inference process, the denoised mel-spectrogram $\tilde{\vm}_0$ can be generated step by step as following:
\begin{equation*}
    \tilde{\vm}_{t-1} = \frac{1}{\sqrt{\alpha_t}}\left( \tilde{\vm}_t - \frac{1-\alpha_t}{\sqrt{1-\bar\alpha_t}} \bepsilon_\theta(\tilde{\vm}_t, \vs, \vc, t) \right) + \sigma_t \bz,
\end{equation*}
where $\bz \sim\mathcal{N}(\bzero, \bI)$ and $t$ decreases gradually from $S$ to 1. Step $S$ can be much smaller than the total step $T$ for the shallow diffusion strategy. In DurIAN-E, $\tilde{\vm}_{S}$ is the actually the output mel-spectrograms from the AR decoder and we set $T=70$ and $S=30$ empirically\footnote{Too big $S$ may lead to the distortion of spectral details and too small $S$ can degrade the denoiser performance.}.
\section{Experiments}
\label{sec:exp}

\begin{table*}[t]
     
   % \small
%\centering

    \begin{tabular}{c| C{1.8cm}c| ccc|C{2.2cm}}
   
       \toprule
       \textbf{System} & \textit{GT} & \textit{GT (Mel + vocoder)} & \textit{DurIAN} &  \textit{FastSpeech 2} &  \textit{DiffSpeech} &\textbf{\textit{DurIAN-E}} \\
          \midrule
        \textbf{MOS} & 4.45 $\pm$ 0.16 & 4.23 $\pm$ 0.17 & 3.73 $\pm$ 0.15 & 3.62 $\pm$ 0.17 & 3.78 $\pm$ 0.19 & \textbf{3.86 $\pm$ 0.14} \\
            \bottomrule
    \end{tabular}
     \label{tab:mos}
    \centering
      \caption{The MOS values of different systems with 95\% confidence intervals.}
       
\end{table*}
\subsection{Experiment setup}
To evaluate the performance of the proposed DurIAN-E model, 
a multi-style Chinese corpus containing 11.8 hours  of  speech data pronounced by 7 different speakers was used as the training dataset. We conducted a set of rich and  fine-grained style tags including ``neural, happy, sad, angry, exciting, annoying, amazing, doubtful, cunning,
solemn, enchanting and taunting". 64 sentences those were not present in the training set were used as the test set. We conducted MOS and ablation preference tests to measure the performance of different systems and different modules by comprehensively assessing the expressiveness and speech quality. For each test,  20 test utterances randomly selected from the test set were synthesized by different systems and evaluated in random order by 10 listeners.\footnote{Examples of synthesized speech by different systems are available at \url{https://sounddemos.github.io/durian-e}.} 
All these systems besides the proposed DurIAN-E shared an unified BigVGAN vocoder \cite{bigvgan} which was trained conditioned on ground truth mel-spectrograms to better compare the performances among different acoustic models.

\subsection{MOS test}
Several state-of-the-art  speech synthesis systems including  \textit{DurIAN} \cite{yu2020durian},  \textit{FastSpeech 2} \cite{renfastspeech} and  \textit{DiffSpeech} \cite{liu2022diffsinger}  were established for comparison and the MOS results are listed in Table 1.  The auto-regressive model \textit{DurIAN}  outperforms the  parallel model \textit{FastSpeech 2}, which
 proves the AR decoder in DurIAN which can make full use of previously generated acoustic features can synthesize better quality speech. Due to the employement of diffusion modules, \textit{DiffSpeech}  achieves better MOS result than \textit{DurIAN} and \textit{FastSpeech 2}.  \textit{DurIAN-E} combines the advantages of those models, which consists of both AR mechanism in  \textit{DurIAN} and the DDPM module in \textit{DiffSpeech} and other complicated structures such as SwishRNNs and SAIN layers.
 Therefore \textit{DurIAN-E} can generate  more expressive and better quality speech and  achieve the best MOS score among all systems,
    which demonstrates  model capacity of the proposed system is sufficient.
\subsection{Ablation test}

\begin{figure}[t]
  \centering
  \includegraphics[width=0.95\linewidth]{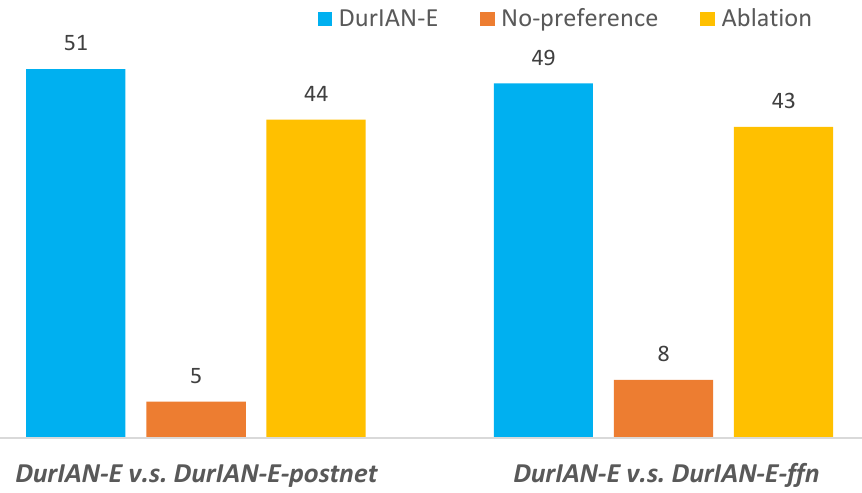}
  \caption{Preference test scores between \textbf{\textit{DurIAN-E}} and ablation systems. The $p$-values of $t$-test are 0.311 and 0.378.}
  \label{fig:abx}
\end{figure}

To demonstrate the effectiveness of using different proposed modules in DurIAN-E, two additional systems were conducted for ablation studies as following:
\begin{itemize}
\setlength{\itemindent}{-5pt}
\setlength{\itemsep}{0pt}
\item \emph{\textbf{DurIAN-E}}: The proposed system as described in Fig. \ref{fig:structure};
\item \emph{\textbf{DurIAN-E-postnet}}: The model using \textit{post-net} \cite{tacotron2} as the denoiser instead of DDPM;
\item \emph{\textbf{DurIAN-E-ffn}}: Using standard Transformers as the linguistic encoder instead of SwishRNN-based ones.
\end{itemize}
The results of two preference tests are shown  in Fig. \ref{fig:abx}. \emph{\textbf{DurIAN-E}} can produce better results than the two ablation systems, which verifies the effectiveness of adopting DDPM denoiser and using SwishRNNs as the linguistic encoder. Comparing with \emph{\textbf{DurIAN-E}}, \textit{post-net} based denoiser rather than the one based on DDPM causes obvious speech quality drop and replacing the SwishRNN blocks with the FFN blocks can results in model stability and pronunciation drop. Meanwhile, the difference between  \emph{\textbf{DurIAN-E}} and  \emph{\textbf{DurIAN-E-postnet}} is bigger than that between    \emph{\textbf{DurIAN-E}} and  \emph{\textbf{DurIAN-E-ffn}}  as depicted from Fig. \ref{fig:abx}, which indicates the effectiveness of the DDPM-based denoiser with SAIN layers  is much more significant.

\section{Conclusion}
\label{sec:con}
In this paper, we propose DurIAN-E, an improved model  for expressive and high-fidelity  TTS, which utilizes SwishRNN-based Transformers as phoneme-level encoder and SAIN-based frame-level encoder to achieve more natural prosody and more expressive speech. A DDPM-based denoiser for mel-spectrograms using SAIN layers is conducted to further improve speech quality and expressiveness. Experimental results of subjective tests prove that DurIAN-E can achieve better performance than the state-of-the-art approaches.
We will increase the DDPM inference efficiency by reducing sample steps and further improve speech quality and accelerate inference speed by incorporating DurIAN-E with other modules such as  conditional variational autoencoder and adversarial learning. We will also deploy the proposed algorithms on online products. 
\bibliographystyle{IEEEbib}
\bibliography{am}

\end{document}

%% file: math_commands.tex
\usepackage{amsmath,amsfonts,bm}

\def\eqref#1{equation~\ref{#1}}
\def\1{\bm{1}}

\def\vb{{\bm{b}}}
\def\vc{{\bm{c}}}

\def\ve{{\bm{e}}}

\def\vh{{\bm{h}}}

\def\vm{{\bm{m}}}

\def\vs{{\bm{s}}}

\def\vx{{\bm{x}}}

\def\mC{{\bm{C}}}

\def\mW{{\bm{W}}}
\def\mX{{\bm{X}}}

\DeclareMathAlphabet{\mathsfit}{\encodingdefault}{\sfdefault}{m}{sl}
\SetMathAlphabet{\mathsfit}{bold}{\encodingdefault}{\sfdefault}{bx}{n}

\newcommand{\E}{\mathbb{E}}

\newcommand{\Eb}[2]{\E_{#1}\!\left[#2\right]}
\newcommand{\bzero}{\mathbf{0}}

\newcommand{\bz}{\mathbf{z}}

\newcommand{\bepsilon}{{\boldsymbol{\epsilon}}}

\newcommand{\bI}{\mathbf{I}}

%% file: draft.bbl
\begin{thebibliography}{10}

\bibitem{tacotron2}
Jonathan Shen, Ruoming Pang, Ron~J Weiss, Mike Schuster, Navdeep Jaitly,
  Zongheng Yang, Zhifeng Chen, Yu~Zhang, Yuxuan Wang, Rj~Skerrv-Ryan, et~al.,
\newblock ``Natural {TTS} synthesis by conditioning {WaveNet} on mel
  spectrogram predictions,''
\newblock in {\em 2018 IEEE international conference on acoustics, speech and
  signal processing (ICASSP)}. IEEE, 2018, pp. 4779--4783.

\bibitem{renfastspeech}
Yi~Ren, Chenxu Hu, Xu~Tan, Tao Qin, Sheng Zhao, Zhou Zhao, and Tie-Yan Liu,
\newblock ``{FastSpeech} 2: Fast and high-quality end-to-end text to speech,''
\newblock in {\em International Conference on Learning Representations}.

\bibitem{paralleltacotron2}
Isaac Elias, Heiga Zen, Jonathan Shen, Yu~Zhang, Ye~Jia, R.J. Skerry-Ryan, and
  Yonghui Wu,
\newblock ``{Parallel Tacotron} 2: A non-autoregressive neural {TTS} model with
  differentiable duration modeling,''
\newblock in {\em Proc. Interspeech 2021}, 2021, pp. 141--145.

\bibitem{yu2020durian}
Chengzhu Yu, Heng Lu, Na~Hu, Meng Yu, Chao Weng, Kun Xu, Peng Liu, Deyi Tuo,
  Shiyin Kang, Guangzhi Lei, et~al.,
\newblock ``Dur{IAN}: Duration informed attention network for speech
  synthesis,''
\newblock 2020.

\bibitem{tits2019visualization}
No{\'e} Tits, Fengna Wang, Kevin El~Haddad, Vincent Pagel, and Thierry Dutoit,
\newblock ``Visualization and interpretation of latent spaces for controlling
  expressive speech synthesis through audio analysis,''
\newblock {\em Proc. Interspeech 2019}, pp. 4475--4479, 2019.

\bibitem{tits2020exploring}
No{\'e} Tits, Kevin El~Haddad, and Thierry Dutoit,
\newblock ``Exploring transfer learning for low resource emotional {TTS},''
\newblock in {\em Intelligent Systems and Applications: Proceedings of the 2019
  Intelligent Systems Conference (IntelliSys) Volume 1}. Springer, 2020, pp.
  52--60.

\bibitem{wang2018style}
Yuxuan Wang, Daisy Stanton, Yu~Zhang, RJ-Skerry Ryan, Eric Battenberg, Joel
  Shor, Ying Xiao, Ye~Jia, Fei Ren, and Rif~A Saurous,
\newblock ``Style tokens: Unsupervised style modeling, control and transfer in
  end-to-end speech synthesis,''
\newblock in {\em International Conference on Machine Learning}. PMLR, 2018,
  pp. 5180--5189.

\bibitem{skerry2018towards}
RJ~Skerry-Ryan, Eric Battenberg, Ying Xiao, Yuxuan Wang, Daisy Stanton, Joel
  Shor, Ron Weiss, Rob Clark, and Rif~A Saurous,
\newblock ``Towards end-to-end prosody transfer for expressive speech synthesis
  with {Tacotron},''
\newblock in {\em international conference on machine learning}. PMLR, 2018,
  pp. 4693--4702.

\bibitem{min2021meta}
Dongchan Min, Dong~Bok Lee, Eunho Yang, and Sung~Ju Hwang,
\newblock ``{Meta-StyleSpeech}: Multi-speaker adaptive text-to-speech
  generation,''
\newblock in {\em International Conference on Machine Learning}. PMLR, 2021,
  pp. 7748--7759.

\bibitem{liu2021meta}
Songxiang Liu, Dan Su, and Dong Yu,
\newblock ``{Meta-Voice}: Fast few-shot style transfer for expressive voice
  cloning using meta learning,''
\newblock {\em arXiv preprint arXiv:2111.07218}, 2021.

\bibitem{li2022styletts}
Yinghao~Aaron Li, Cong Han, and Nima Mesgarani,
\newblock ``{StyleTTS}: A style-based generative model for natural and diverse
  text-to-speech synthesis,''
\newblock {\em arXiv preprint arXiv:2205.15439}, 2022.

\bibitem{lei2022simple}
Tao Lei, Ran Tian, Jasmijn Bastings, and Ankur~P Parikh,
\newblock ``Simple recurrence improves masked language models,''
\newblock {\em arXiv preprint arXiv:2205.11588}, 2022.

\bibitem{ddpm}
Jonathan Ho, Ajay Jain, and Pieter Abbeel,
\newblock ``Denoising diffusion probabilistic models,''
\newblock {\em Advances in Neural Information Processing Systems}, vol. 33, pp.
  6840--6851, 2020.

\bibitem{jeong2021diff}
Myeonghun Jeong, Hyeongju Kim, Sung~Jun Cheon, Byoung~Jin Choi, and Nam~Soo
  Kim,
\newblock ``Diff-{TTS}: A denoising diffusion model for text-to-speech,''
\newblock {\em arXiv preprint arXiv:2104.01409}, 2021.

\bibitem{liu2022diffsinger}
Jinglin Liu, Chengxi Li, Yi~Ren, Feiyang Chen, and Zhou Zhao,
\newblock ``Diff{S}inger: Singing voice synthesis via shallow diffusion
  mechanism,''
\newblock in {\em Proceedings of the AAAI Conference on Artificial
  Intelligence}, 2022, vol.~36, pp. 11020--11028.

\bibitem{transformer}
Ashish Vaswani, Noam Shazeer, Niki Parmar, Jakob Uszkoreit, Llion Jones,
  Aidan~N Gomez, {\L}ukasz Kaiser, and Illia Polosukhin,
\newblock ``Attention is all you need,''
\newblock {\em Advances in neural information processing systems}, vol. 30,
  2017.

\bibitem{ramachandran2017searching}
Prajit Ramachandran, Barret Zoph, and Quoc~V Le,
\newblock ``Searching for activation functions,''
\newblock {\em arXiv preprint arXiv:1710.05941}, 2017.

\bibitem{diffwave}
Zhifeng Kong, Wei Ping, Jiaji Huang, Kexin Zhao, and Bryan Catanzaro,
\newblock ``Diffwave: A versatile diffusion model for audio synthesis,''
\newblock {\em arXiv preprint arXiv:2009.09761}, 2020.

\bibitem{wavenet}
A{\"a}ron van~den Oord, Sander Dieleman, Heiga Zen, Karen Simonyan, Oriol
  Vinyals, Alex Graves, Nal Kalchbrenner, Andrew Senior, and Koray Kavukcuoglu,
\newblock ``{WaveNet}: A generative model for raw audio,''
\newblock in {\em 9th ISCA Speech Synthesis Workshop}, pp. 125--125.

\bibitem{bigvgan}
Sang-gil Lee, Wei Ping, Boris Ginsburg, Bryan Catanzaro, and Sungroh Yoon,
\newblock ``{BigVGAN}: A universal neural vocoder with large-scale training,''
\newblock in {\em The Eleventh International Conference on Learning
  Representations}, 2022.

\end{thebibliography}
